# Evaluation of Setup Uncertainties for Single-Fraction SRS by Comparing the Two Different Mask-Creation Methods

**Jong Geun Baek**

*Department of Physics, Yeungnam University, Gyeongsan 712-749 and*

*Department of Radiation Oncology, Dongguk University Gyeongju Hospital 780-350, Korea*

**Hyun Soo Jang**

*Department of Radiation Oncology, Dongguk University Gyeongju Hospital 780-350, Korea*

**Young Kee Oh**

*Department of Radiation Oncology, Keimyung University Dongsan Medical Center, Daegu 700-712, Korea*

**Hyun Jeong Lee, Eng Chan Kim[*]**

*Department of Physics, Yeungnam University, Gyeongsan 712-749, Korea*

# ABSTRACT

The purpose of this study was to evaluate the setup uncertainties for single-fraction stereotactic radiosurgery (SF-SRS) based on the clinical data with the two different mask-creation methods using pretreatment CBCT imaging guidance. Dedicated frameless fixation BrainLAB masks for 23 patients were created as a routine mask (R-mask) making method, as explained in the BrainLAB's user manual. The alternative masks (A-mask) which were created by modifying the cover range of the R-mask for the patient's head were used for 23 patients. The systematic errors including the each mask and stereotactic target localizer were analyzed and the errors were calculated as the mean ± standard deviation (SD) from the LR, SI, AP, and yaw setup corrections. In addition, the frequency of three-dimensional (3D) vector length were also analyzed. The values of the mean setup corrections for the R-mask in all directions were small; < 0.7 mm and < 0.1°, whereas the magnitudes of the SDs were relatively large compared to the mean values. In contrast to the R-mask, the means and SDs of the A-mask were smaller than those for the R-mask with the exception of the SD in the AP direction. The mean and SD in the yaw rotational direction in the R-mask and A-mask system were comparable. The 3D vector shifts of a larger magnitude occurred more frequently for the R-mask than the A-mask. The setup uncertainties for each mask with the stereotactic localizing system had an asymmetric offset towards the positive AP direction. The A-mask-creation method, which is capable of covering the top of the patient's head is superior to that for R-mask, and thereby the use of the A-mask is encouraged for SF-SRS to reduce the setup uncertainties. Moreover, the careful mask making is required to prevent the possible setup uncertainties.

Email: eckim@yu.ac.kr

Fax: +82-53-810-2343

# I. INTRODUCTION

The development of image-guided systems has led to the increasingly popularity of single-fraction stereotactic radiosurgery (SF-SRS) for the treatment of primary brain tumors and solitary brain metastases [1]. In addition, SF-SRS has no inter-fraction patient random error and machine systematic error [2-3]. On the other hand, SF-SRS can lead to an increased risk of complications if the geometric accuracy is insufficient because of its high prescription dose in a single fraction. Therefore, high geometric accuracy is particularly important for the clinical applications of SF-SRS [4]. A number of studies have verified the geometric uncertainty of patient immobilization and localization systems, such as invasive frame [5-8] or frameless-based dedicated mask devices [7-11], and conventional thermoplastic mask with or without a dental bite block [12-14] using a range of image-guided systems. Hong et al. [6] reported that the setup corrections in the translational directions were within 1 mm using kilovoltage orthogonal images with an onboard imager (OBI). Ramakrishna et al. [7] reported that the overall system accuracy of the frameless technique with BrainLAB ExacTrac stereoscopic X-ray system is comparable to that of the frame-based approach. Guckenberger et al. [15] reported that the overall setup error could be reduced from $3.9 \pm 1.7$ mm to $0.9 \pm 0.6$ mm by cone beam computed tomography (CBCT). As seen in the previous findings, the setup uncertainties could be reduced considerably through the use of advanced image-guided systems. On the other hand, residual systematic errors, which can contribute to setup uncertainties, still exist for a variety of reasons and the magnitudes of the errors differ from institution to institution. Therefore, each institution where SF-SRS can be performed needs to evaluate the institution's specific systematic uncertainties and to make every effort to minimize the setup uncertainties before initiating treatment. The author's institution have attempted to minimize the setup uncertainties of SF-SRS in many ways, such as the modified mask creation method, locking the automatic couch movement, and quality assurance for the alignment of each isocenter used for treatment. The present study evaluated the setup uncertainties for SF-SRS based on the clinical data with the two different mask-creation methods using pretreatment CBCT imaging guidance, and compared the setup uncertainties with the previous findings.

## II. MATERIALS AND METHODS

**1. Selected patients and two frameless mask-creation methods**

Between January 2011 and March 2014, 46 patients were treated with SF-SRS immobilized with a BrainLAB frameless thermoplastic head mask at the author's institution. From January 2011 to August 2012, as shown in Fig. 1 (a), dedicated masks for 23 patients were created as a routine mask (R-mask) making method, as explained in the BrainLAB's user manual. However, the R-mask only covered the lower region of the patient frontal bone. Since September 2012, the R-mask making method was replaced with an alternative mask (A-mask), which was a modified mask-creation method in an expanded shape for 23 patients. The A-mask was capable of covering the top of the patient's head, as shown in Fig.1 (b).

**2. CT image acquisition and planning for SF-SRS**

For SF-SRS treatment planning, all patients underwent CT using a Somatom Emotion 6 helical CT scanner (Siemens Medical Systems, Erlangen, Germany) with a 2 mm slice thickness. During CT image acquisition, the BrainLAB localizer was mounted onto a dedicated frameless mask system to provide three-dimensional (3D) stereotactic coordinates for target localization. The resulting CT images were transferred to iPlan planning system 3.0.2 (BrainLAB, Feldkirchen, Germany). The gross tumor volume of each patient was contoured and based on the magnetic resonance imaging (MRI) scans by comparing the planning CT images. At least 6 or more non-coplanar arc beams were planned by trained medical physicist.

**3. Quality assurance for geometric accuracy of SF-SRS**

To assess the setup uncertainty of each mask more accurately, other factors that can affect the systematic uncertainty in SF-SRS should be reduced as much as possible. These factors include the misalignment of room lasers with the radiation isocenter and discrepancies between the CBCT isocenter and treatment machine radiation isocenter. Quality assurance (QA) for the accuracy of the laser alignment to the radiation isocenter was carried out using the Winston-Lutz test with a tolerance level of 0.5 mm for a different collimators, couches and gantry angles. The accuracy of the CBCT isocenter to the treatment machine isocenter was checked using a standard OBI QA phantom with a threshold within ± 0.5 mm in the left-right (LR), superior-inferior (SI), anterior-posterior (AP). This QA procedure was performed immediately before the patient's treatment.

**4. Patient setup, CBCT image acquisition and registration**

All SF-SRS patients were immobilized using a frameless head mask made by each creation method. The patients were then positioned and aligned according to the stereotactic BrainLAB target positioner using in-room setup lasers. Subsequently, the CBCT images were acquired using the gantry-mounted OBI systems in low-dose head mode in the available scanning protocol. The CBCT imaging acquisition time was approximately 30 seconds for the mode with a gantry rotation of 200°. All acquired CBCT images had a sufficient scan length to register between the CBCT images and planning CT images with a 2 mm slice thickness. The 3D image registration was first performed automatically using the OBI registration system and was then checked manually by therapists based on the bony anatomy and target region. The image registrations then were reviewed and confirmed by the treating physicians to ensure that the matching was accurate. After 3D image registration, the OBI software provided translational and rotational deviations, as suggested by the automatic adjustment couch shifts LR, SI, AP and yaw directions to correct the patient setup errors.

**5. Clinical data analysis and statistical considerations**

The setup correction data for SF-SRS patient in all directions was collected into registration software as an Offline Review (Varian Medical System). Clinical data analysis consisted of three approaches to evaluate the institution's specific systematic uncertainties. The systematic error of each mask, the stereotactic target localizing systematic error, and the frequency of the 3D vector length were analyzed. The systematic uncertainties of each mask and the localizing system were calculated as the mean ± standard deviation (SD) from the LR, SI, AP, and yaw corrections. Box-and-whisker plots were used to represent the distribution of the setup corrections for each mask system in all directions. The overall distributions of the setup corrections in all directions for each mask system were plotted to assess the stereotactic localizing systematic uncertainties. The frequencies of the 3D vector displacements were quantified to determine how often translational corrections occurred for the target localizing system, and the results were calculated using the following formula: $\sqrt{x^2 + y^2 + z^2}$, where x, y and z represent the LR, SI and AP setup correction, respectively. A t test using PASW statistics 18.0 (SPSS Inc., Chicago, IL, USA) was used to determine if a statistically significant difference was present between the R-mask and A-mask. The differences were considered significant for $p < 0.05$.

## III. RESULTS

Table. 1 lists the mean setup corrections for each mask in all directions. The values for the R-mask in all directions were small; < 0.7 mm and < 0.1°. On the other hand, the magnitudes of the SDs were relatively large compared to the mean values. The highest SD was 1.88 mm in the SI direction for the R-mask, as listed in Table. 2. These findings suggest that the R-mask has a factor that may lead to systematic uncertainties in the SI direction. In contrast to the R-mask, the means and SDs of the A-mask were smaller than those for the R-mask with the exception of the SD in the AP direction, as listed in Tables 1-2. The SD in the SI direction decreased from 1.88 mm to 1.14 mm. The mean and SD in the yaw direction in the R-mask and A-mask system were comparable. As shown in Fig.2, the box-and-whisker plots of the setup corrections in all directions for the R-mask were similar to those for the A-mask with the exception of the SI direction. The distribution of setup corrections in the SI direction for A-mask decreased in a positive direction. This suggests that the A-mask can reduce the setup uncertainties in the positive SI direction. Fig. 3 (a) present the cumulative frequencies of the 3D vector distance in the translational setup corrections for each mask. The 3D vector shifts of a larger magnitude occurred more frequently for the R-mask than the A-mask. The 3D vector length of ≥ 1.5 mm was 82.6 % for the R-mask, whereas the distance was only 26.1 % for the A-mask. In particular, the distributions of the 3D vector length for the A-mask decreased rapidly from a starting length of 1 mm, but these variations for the R-mask declined only gradually. Moreover, as shown in Table. 3, statistical analysis indicated that the frequencies of the 3D vector shift in the translational direction for the A-mask were significantly different for the R-mask ($p < 0.05$). Cumulative frequency analysis was also performed on the magnitudes of the yaw direction. In contrast to the 3D vector length of the translational shift, these frequencies were similar regardless of the mask-creation methods, as shown in Table 3 and Fig. 3 (b). Fig. 5 shows the overall distributions of the setup corrections in all directions for each mask with the stereotactic localizing system. The distribution curves for the R-mask in the LR and AP directions had an asymmetric offset towards the positive direction. The biased distribution curve in the positive LR direction was not observed for the A-mask, whereas an asymmetric distribution curve was still represented in the positive AP direction. This result indicated that there were the institution's specific systematic uncertainties of the stereotactic localizing system in the positive AP direction.

## IV. DISCUSSION

The setup uncertainties for SF-SRS can be divided into two main sources of error: patient random error and machine systematic error. The patient random error, which is the intra-fractional variation of each patient, was not analyzed in this study because previous studies using the BrainLAB frameless mask reported that this variation was very small [9, 14]. In addition, the patient random error during treatment depends on each patient's condition and on how well the patient is instructed in the treatment workflow. To minimize the intra-fractional random error in this institution, each patient was given detailed explanations of the adjustable couch movements for target positioning, couch rotations for the noncoplanar arcs, beam on sound during irradiation, and the approximate total treatment time. In addition, the couch was not moved automatically to apply the shifts by the image registration of CBCT and reference planning CT because an abrupt couch adjustment might lead to involuntary patient movement. To prevent movement during the treatment, all the couch movements required for SF-SRS were performed manually by radiation therapists in the treatment room. As a result, the intra-fractional patient random error was quite small, <0.4±0.25 mm in the translational directions, using post-treatment CBCT imaging for 10 patients. Therefore, the intra-fractional patient random error can be reduced significantly by establishing setup protocols but the error was very small.

The systematic errors are another factors that can contribute to the setup uncertainties for SF-SRS. These errors in patient setup result mainly from different sources of errors, which include the systematic uncertainties of fixation mask, systematic uncertainties of stereotactic target positioning system, CBCT systematic uncertainties, and alignment of room lasers with the treatment radiation isocenter [18].

The CBCT systematic uncertainties were unable to be measured by an image guided procedure and the independent test for CBCT should be performed before treatment on the same day. The accuracy of the CBCT isocenter to the radiation isocenter was checked with a threshold within ± 0.5 mm in the LR,

SI and AP directions prior to treatment. Although, this QA procedure can increase the workload, the test should be performed to reduce the image registration uncertainties between the CBCT for the patient setup and planning CT for reference images.

The systematic uncertainties from laser alignment to the radiation isocenter were measured using a Winston-Lutz test with a tolerance of 0.5 mm, as explained in MATERIALS AND METHODS. The systematic uncertainties of CBCT and laser alignment were subtracted from the evaluation to obtain the net systematic errors for the two different fixation masks and stereotactic target positioner.

As shown in Fig. 1 (A), the R-mask was created routinely based on the BrainLAB's user manual until August 2012, and many institutions with the fixation mask system for the treatment of SRS have adopted the R-mask-making method (as shown in figures of references) [7,16-20]. On the other hand, the R-mask has the systematic uncertainties in the SI direction after analyzing the patient setup corrections in Table 2 and Fig. 2. To reduce the systematic uncertainties for the R-mask, the A-mask-making method for patient head fixation have been adopted, as shown in Fig. 1 (B). As a result, the means and SDs in the LR and SI directions were reduced, as shown in Table. 1-2. In particular, the systematic uncertainties in the positive SI direction decreased for the A-mask, as shown in Fig. 2. A possible explanation for this could be that patients using the R-mask might be able to move more easily along the SI direction than patients with the A-mask because the R-mask is not capable of covering the top of the patient's head. In addition, the cumulative frequencies of the 3D vector distance of the translational setup corrections show that the overall setup uncertainties for the R-mask was reduced from 2.17±0.73 to 1.44±0.72 by changing the mask-creation method to the A-mask, and that the differences between the two masks are statistically significant ($p < 0.05$. Based on these results, it was assumed that the A-mask-creation method is more suitable for SF-SRS than that for the R-mask to reduce the systematic uncertainties of translational directions. As shown in Fig. 3 (b), similar behavior was not observed for the yaw rotational direction and the setup errors of the direction in the R-mask and A-mask were similar ($p = 0.92$). A possible explanation for this could be that the magnitudes of the setup errors for the yaw rotational direction were originally very small and the A-mask does not improve the setup accuracy for the yaw rotational direction.

As shown in Fig. 4, the distribution curves in the AP direction for both masks showed an asymmetric offset towards the positive direction. Two factors might explain this finding. First, the stereotactic target positioning system in this institution could have a systematic error of approximately 0.5-1 mm in the positive AP direction. A systematic error can occur when four sheets of paper, which shows the projection on the stereotactic localizer box walls of the treatment isocenter position, are attached incorrectly to the BrainLAB target positioner as a reference for correct beam positioning. Therefore, slight misalignment between the papers and the target positioner could lead to specific systematic errors in this institution for the AP direction. Second, tightening or loosening of the fixation mask due to complete hardening of the mask after the CT simulation might contribute to the setup uncertainties for the AP direction. Based on the BrainLAB's user manual for making the mask, the time required for only top mask hardening is at least 30 minutes. On the other hand, the total time needed for mask creation is approximately 90 minutes. Prolonged mask creation time may cause patient discomfort and the mask could be created improperly to the patient head contour. For this reason, it is difficult to fulfill the mask making time in the BrainLAB's user manual and sometimes, the mask is hardened completely after CT simulation. Therefore, tightening or loosening of the fixation mask at the time of treatment can affect the setup uncertainties for the AP direction.

Table. 4 summarizes the previous findings regarding the setup uncertainties of different fixation mask devices using an image-guided system for SRS. Although these studies are difficult to compare due to the different verification methods and measurement deviations, many studies reported large deviations of the setup uncertainties in the SI direction, even the use of dental fixation. Therefore, setup errors in the SI direction can occur frequently for SRS patients and as a result, each institution needs to make every effort to reduce the setup uncertainties in the SI direction.

In conclusion, based on the results, the A-mask creation method is superior to that for R-mask, and thereby the use of the A-mask is encouraged for SF-SRS to reduce the setup uncertainties. Moreover, the careful mask making is required to prevent the possible setup uncertainties.

## V. CONCLUSION

This study evaluated the setup uncertainties based on the analyzed clinical data of SF-SRS by comparing the two different mask-creation methods using pretreatment CBCT imaging guidance. The A-mask-creation method, which was a modified shape of the R-mask, can reduce the systematic uncertainties of BrainLAB fixation mask system by covering the top of the patient head. Therefore, the A-mask creation method is encouraged for SF-SRS to reduce the setup uncertainties.

Table 1. Mean values of the setup corrections for each mask with stereotactic localizing system in all directions, all magnitudes of means for the A-mask were smaller than those of the R-mask.

| Dedicated mask creation methods | LR (mm) | SI (mm) | AP (mm) | Yaw (°) |
|---|---|---|---|---|
| R-mask | 0.30 | 0.22 | 0.61 | -0.08 |
| A-mask | 0.13 | 0.13 | 0.48 | -0.07 |

Table 2. Standard deviations of each mask in all directions, all magnitudes of SDs for the A-mask were smaller than those for the R-mask with the exception of the AP direction.

| Dedicated mask creation methods | LR (mm) | SI (mm) | AP (mm) | Yaw (°) |
|---|---|---|---|---|
| R-mask | 1.02 | 1.88 | 0.58 | 0.59 |
| A-mask | 0.69 | 1.14 | 0.79 | 0.53 |

Table 3. The mean ± standard deviation of 3D vector lengths for translational setup errors and of yaw rotational setup errors

| | R-mask | A-mask | $p$-value |
|---|---|---|---|
| 3D vector Length (mm) | 2.17±0.73 | 1.44±0.72 | 0.01 |
| Rotational correction (°) | -0.08±0.59 | -0.07±0.53 | 0.92 |

Table.4. Literature overview of the translational setup uncertainties of different fixation mask and image-guided systems

| fixation mask | Image-guided system | Translational setup error (mean ± S.D) | | | | Author |
|---|---|---|---|---|---|---|
| | | 3D Vector [mm] | LR [mm] | SI [mm] | AP [mm] | |
| Thermoplastic mask with bite block | Serial CT scans | 2.2±1.1 | -0.04±1.4 | 0.6±1.8 | -0.1±0.8 | Baumert et al. [12] 2005 |
| BrainLab mask with bite block | CBCT | 3.1±2.1 | 0.5±1.6 | 0.4±2.7 | 0.4±1.9 | Ingrosso et al. [17] 2012 |
| BrainLab mask | kV-OBI (2D) | 2.3±1.0 | 0.5±0.7 | -1.0±1.6 | -0.7±1.2 | Ali et al. [18] 2010 |
| | Serial CT scans | 1.70±0.83 | -0.43±1.49 | 0.55±0.96 | -0.25±0.62 | Theelen et al. [16] 2011 |
| R-mask | CBCT | 2.17±0.73 | 0.3±1.02 | 0.22±1.88 | 0.61±0.58 | Present study |
| A-mask | | 1.44±0.72 | 0.13±0.69 | 0.13±1.14 | 0.48±0.79 | |

Figure Captions.

Fig. 1. (a) Routine mask (R-mask): Dedicated mask created using the routine method based on the BrainLAB's user manual. (b) Alternative mask (A-mask): Dedicated mask created using the modified method which is able to cover the top of the patient head.

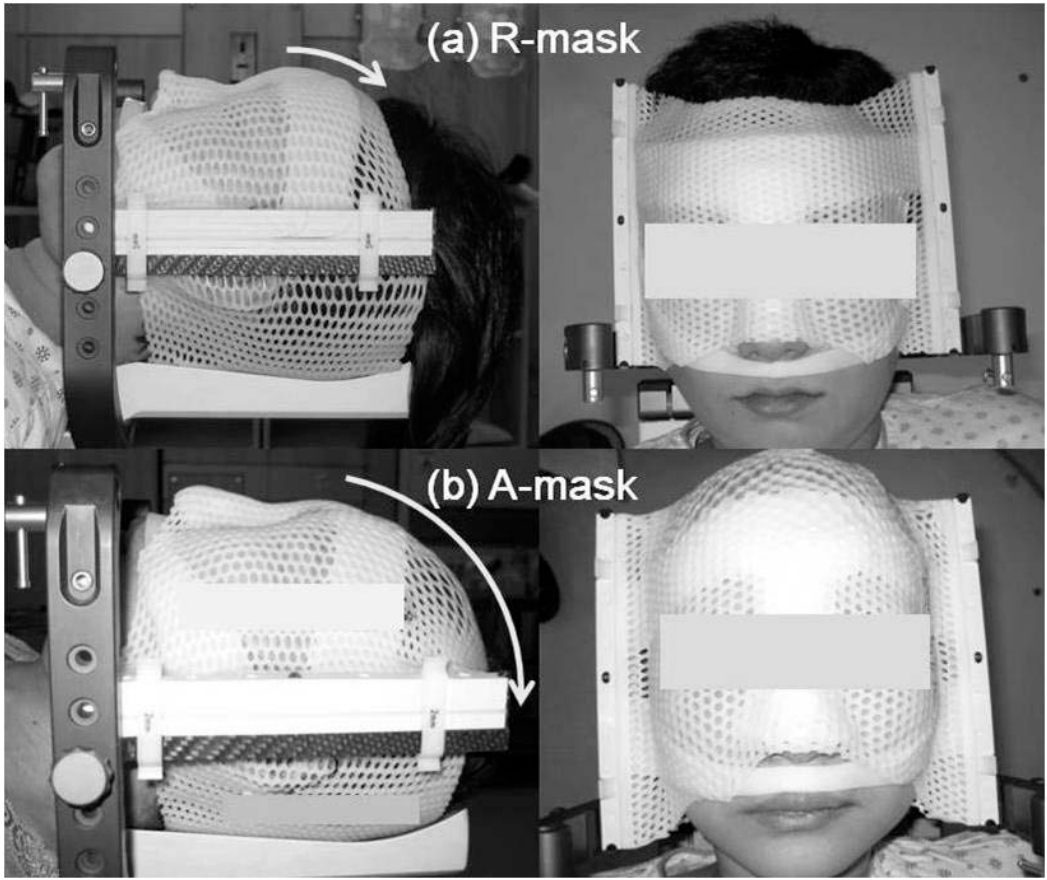

Fig. 2. Comparison for the ranges of setup corrections of each mask in all directions. On each box, the central small circle is the mean, the central bar is the median, the edges of the box are the 25$^{th}$ and 75$^{th}$ percentiles, the whiskers corresponds to approximately ± 1.5 SD, the outliers are plotted as the crosses individually, and the outliers correspond to maximum and minimum observations.

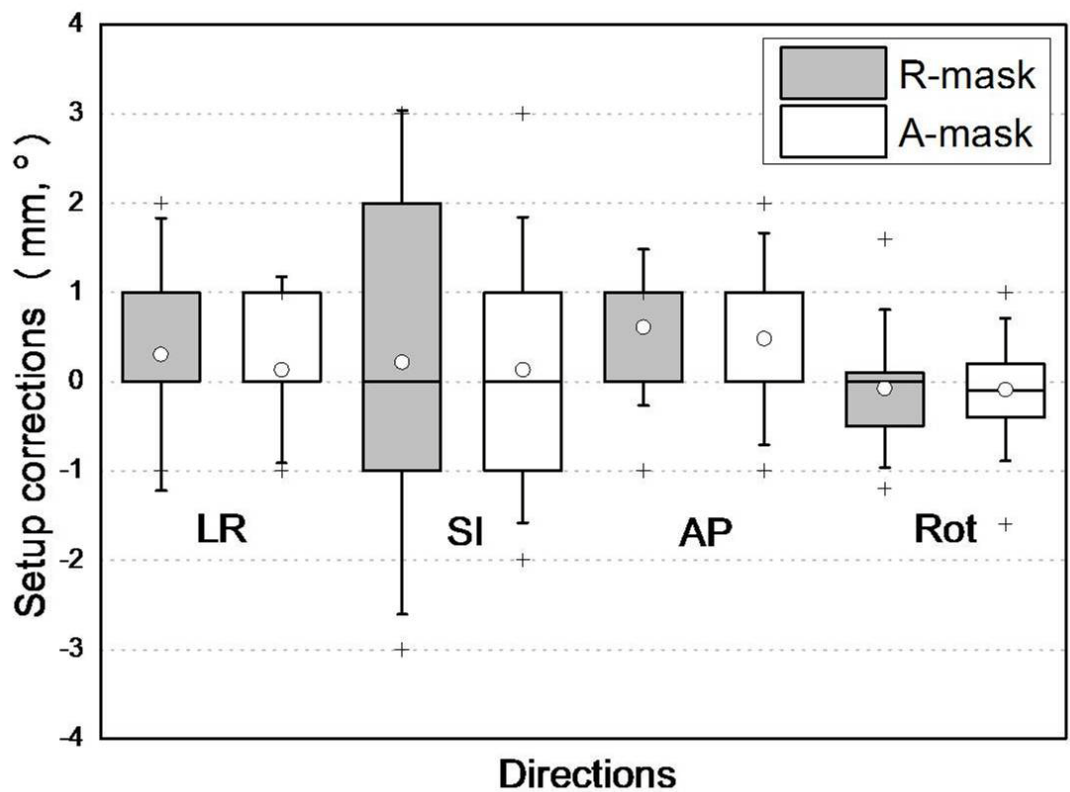

Fig. 3. Cumulative frequencies of translational and rotational setup corrections of each mask with a stereotactic localizing system. (a) The 3D vector distance for the translational setup corrections, a statistically significant difference was observed between R-mask and A-mask ($P < 0.05$) (b) The rotational setup corrections, no statistically significant difference was observed between the two masks.

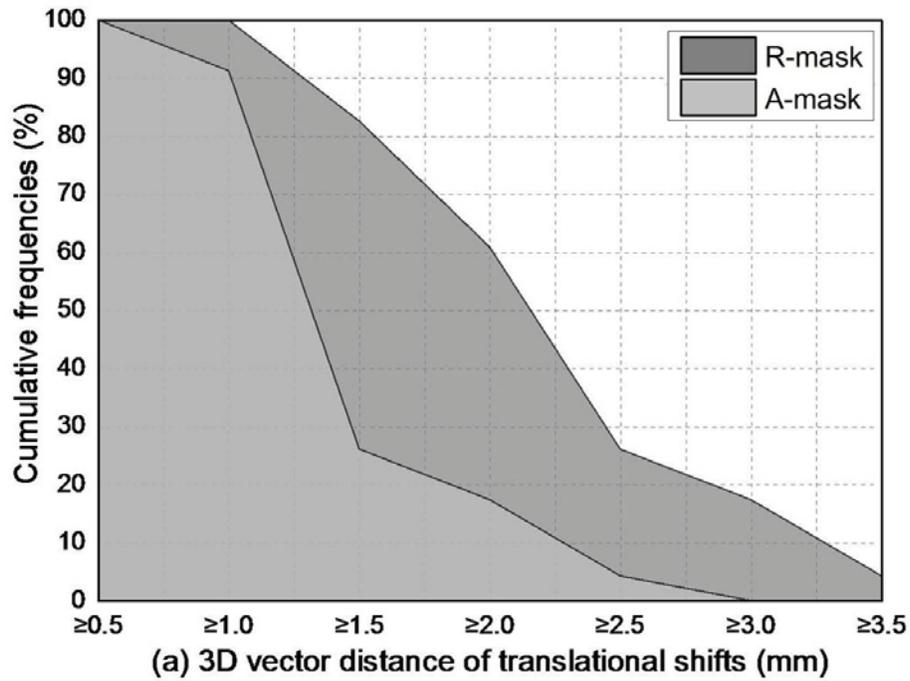

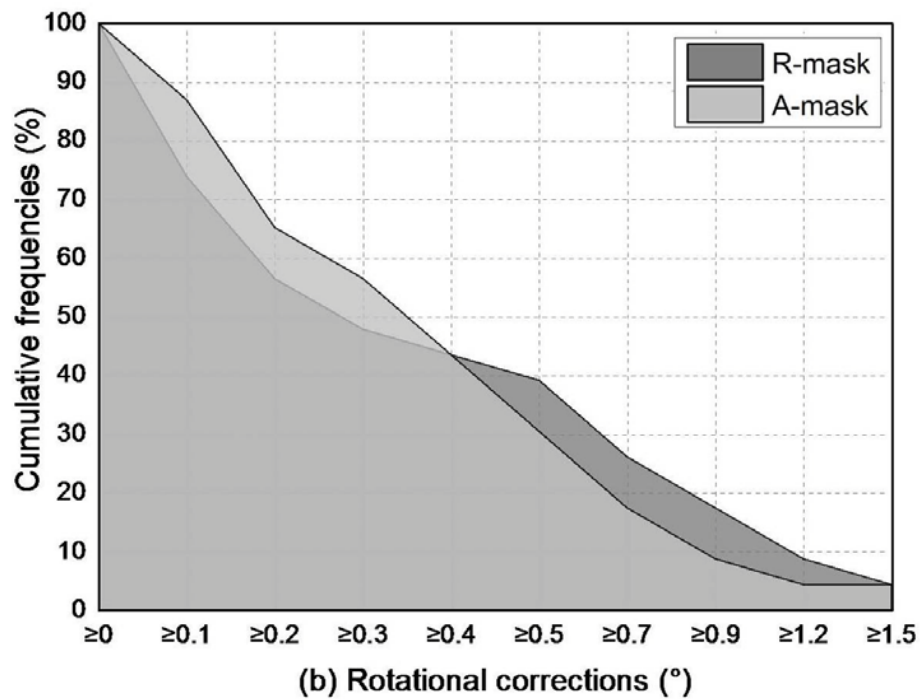

Fig. 4. Overall distributions of the setup corrections in all directions for each mask with a stereotactic localizing system. The distribution for the R-mask in the SI direction was broadest and the distributions for both mask systems in the AP direction represent an asymmetric offset toward the positive direction.

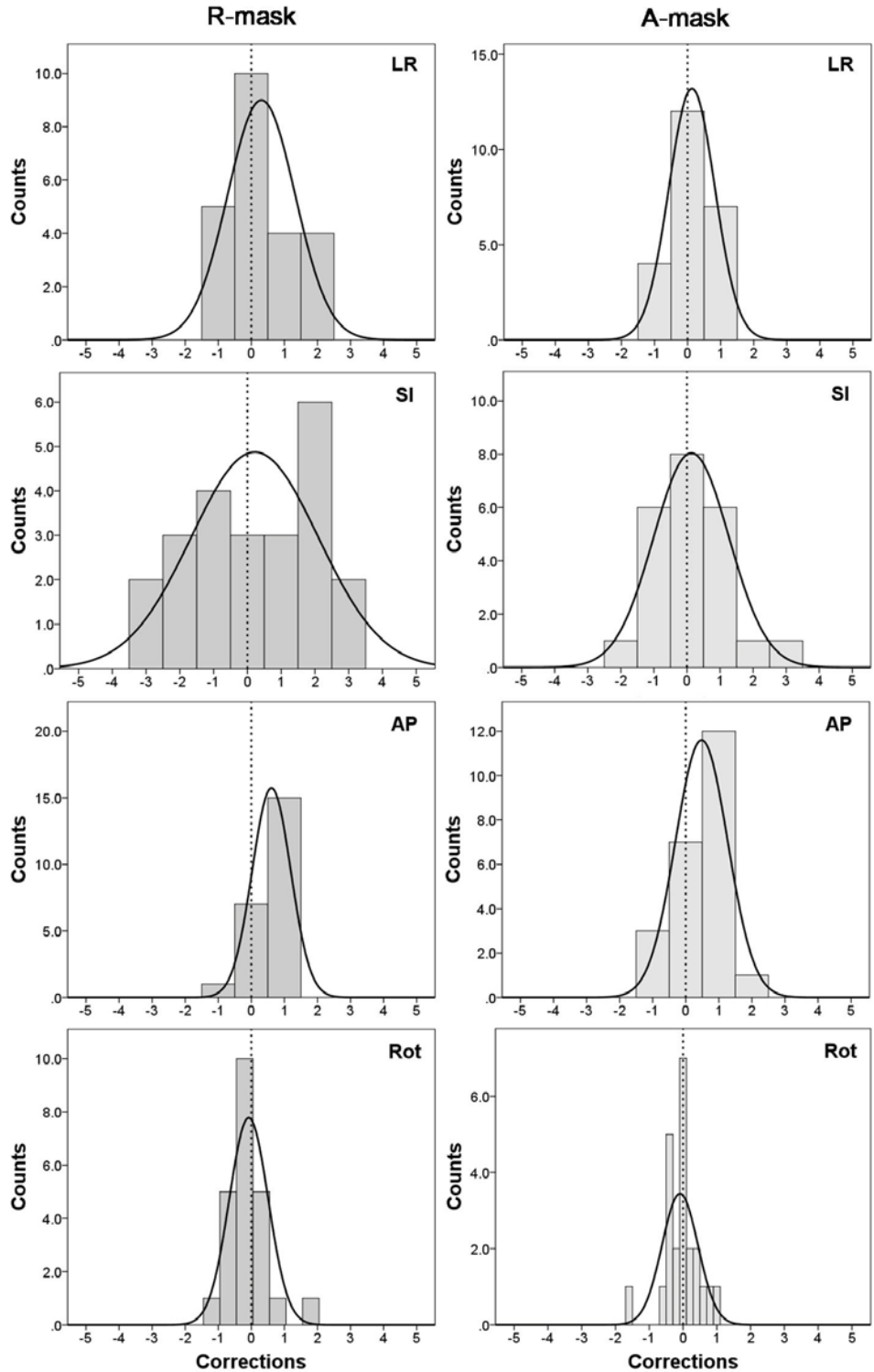